\documentclass[12pt]{article}
\usepackage[english]{babel}
\usepackage[latin1]{inputenc}
\usepackage{amsfonts,amssymb,amsmath, epsfig}
\usepackage{color,graphicx,graphics,psfrag}
\usepackage{amsmath,amstext,amssymb,amsfonts, amscd}

\usepackage{hyperref}           

\def\Box{\leavevmode\vbox{\hrule
     \hbox{\vrule\kern4pt\vbox{\kern4pt}%
           \vrule}\hrule}}
\def\blackbox{\leavevmode\vrule height 5pt width 4pt depth 0pt\relax}
\def\endproof{\null\hfill {$\blackbox$}\bigskip}

\newcounter{appendix}
\def\appendix{\advance\c@appendix by 1
   \def\thesection{\Alph{section}}
   \ifnum\c@appendix=1 \setcounter{section}{-1} \fi
   \@startsection {section}{1}{\z@}{-3.5ex plus -1ex minus
   -.2ex}{2.3ex plus .2ex}{\Large\bf}}

\newcommand{\si}{\sigma_R}
\newcommand{\s}{\sigma}
\newcommand{\bb}{\begin{equation}}
\newcommand{\ee}{\end{equation}}
\newcommand{\bq}{\begin{eqnarray}}
\newcommand{\eq}{\end{eqnarray}}
\newcommand{\bqn}{\begin{eqnarray*}}
\newcommand{\eqn}{\end{eqnarray*}}
\newcommand{\n}{\nonumber}

\def\paragraph#1{{\bf #1\ }}

\newtheorem{lemma}{Lemma}[section]

\newtheorem{theorem}[lemma]{Theorem}

\newtheorem{proposition}[lemma]{Proposition}

\newtheorem{remark}{Remark}[section]

\title{Well-posedness for Hall-magnetohydrodynamics}
\author{Dongho Chae $^{(1)}$, Pierre Degond$^{(2,3)}$, Jian-Guo Liu$^{(4)}$}
\date{}
\begin{document}

\maketitle

\vspace{0.5 cm}

\begin{center}
1-Department of Mathematics;\\
 Chung-Ang
University\\
 Seoul 156-756, Korea\\
  e-mail: { dchae@cau.ac.kr }\\
\end{center}
\begin{center}
2-Université de Toulouse; UPS, INSA, UT1, UTM ;\\
Institut de Mathématiques de Toulouse ; \\
F-31062 Toulouse, France. \\
3-CNRS; Institut de Mathématiques de Toulouse UMR 5219 ;\\
F-31062 Toulouse, France.\\
email:  pierre.degond@math.univ-toulouse.fr
\end{center}
\begin{center}
4- Department of Physics and Department of Mathematics\\
Duke University\\
Durham, NC 27708, USA\\
email: jliu@phy.duke.edu
\end{center}

\vspace{0.5 cm}
\begin{abstract}
We prove local existence of smooth solutions for large data and global smooth solutions for small data to the incompressible, resitive, viscous or inviscid Hall-MHD model. We also show a Liouville theorem for the stationary solutions.
\end{abstract}

\medskip
\noindent
{\bf Acknowledgements:} The first and third authors wish to thank the hospitality of the
 Institut de Math\'ematiques de Toulouse, France, where this research has been carried through,
  under funding from the Centre National de la Recherche Scientifique. The research of the first author is also
  supported partially by  NRF Grant no. 2006-0093854,
and by the Chung-Ang University Research Grants in 2012.
   The second authors acknowledges
   support by the french 'Agence Nationale pour la Recherche (ANR)' in the frame of the contract
'BOOST' (ANR Blanc 2010).

\medskip
\noindent {\bf Key words:} Hall-MHD, smooth solutions, well-posedness, Liouville theorem

\medskip
\noindent
{\bf AMS Subject classification: } 35L60, 35K55, 35Q80
\vskip 0.4cm

\setcounter{equation}{0}
\section{Introduction}
\label{sec:intro}

In this paper, we study the existence of smooth solutions for the incompressible resistive Hall-MagnetoHydroDynamics system (in short, Hall-MHD). While usual incompressible resistive MHD equations are well understood for quite long time (see e.g. \cite{Duvaut_Lions_ARMA72}), Hall-MHD has received little attention from mathematicians. However, in many current physics problems, Hall-MHD is required. The first systematic study of Hall-MHD is due to Lighthill \cite{Lighthill_PhilTransRSocLond60} followed by
Campos \cite{Campos_TheoretComputFluidDynamic98}. The Hall-MHD is indeed needed for such problems as magnetic reconnection in space plasmas \cite{Forbes_GeoAstroFluidDyn91, Homann_Grauer_PhysicaD05}, star formation \cite{Wardle_AstroSpaceSci04, Balbus_Terquem_AstroJ01}, neutron stars \cite{Shalybkov_Urpin_AstronAstrophys97} and geo-dynamo \cite{Mininni_etal_AstroJ03}. A physical review on these questions can be found in \cite{Polygiannakis_Moussas_PlasmaPhysControlFusion01}. Mathematical derivations of Hall-MHD equations from either two-fluids or kinetic models can be found in \cite{Acheritogaray_etal_KRM11} and in this paper, the first existence result of global weak solutions is given. In \cite{Charles_etal_KRM13}, a stability analysis of a Vlasov equation modeling the Hall effect in plasmas is carried over.

Hall-MHD is believed to be an essential feature in the problem of magnetic reconnection. Magnetic reconnection corresponds to changes in the topology of magnetic field lines which are ubiquitously observed in space. However, in ideal MHD, due to ideal Ohm's law, the magnetic field undergoes a passive transport by the fluid velocity and its topology is preserved. The Hall term restores the influence of the electric current in the Lorenz force occuring in Ohm's law, which was neglected in conventional MHD. This term is quadratic in the magnetic field and involves second order derivatives. So its influence becomes dominant in the cases where the magnetic shear is large, which precisely occurs during reconnection events. In laminar situations, this term is usually small and can be neglected, which is why conventional MHD models ignore it.

In this paper, we focus on the mathematical analysis of this model and investigate the existence and uniqueness of smooth solutions. We also prove a Liouville theorem for stationary solutions. The main results are stated in section \ref{sec:main}. Theorem \ref{thm:global_weak} provides the global existence of weak solutions for any data. Compared to \cite{Acheritogaray_etal_KRM11} which dealt with a periodic setting, the present result concerns the whole space case. However, the proof is identical and is omitted. Theorem \ref{thm:blowup} shows the local existence of smooth solutions for large data and provides a blow-up criterion. Theorem \ref{thm:global_small} proves the global existence of smooth solutions for small data. Theorem \ref{thm:unique} gives the uniqueness of the solution. Finally, a Liouville theorem for stationary solutions is provided in Theorem \ref{thm:Liouville}. The main technical point is to control the second order derivatives in the Hall term by the diffusion term induced by the resistivity. This can be done thanks to the special antisymmetric structure of the Hall term. The proofs are carried over in section \ref{sec:proof}.

\setcounter{equation}{0}
\section{Statement of the main results}
\label{sec:main}

We consider the following viscous or invisicd,  resistivity incompressible MHD-Hall equations.
\begin{eqnarray}
& & \hspace{-1cm}  \partial_t u +  u \cdot \nabla u + \nabla p  =  (\nabla \times B) \times B + \nu\Delta u,  \label{eq:dtu} \\
& & \hspace{-1cm} \nabla \cdot u = 0 , \label{eq:divu}\\
& & \hspace{-1cm} \partial_t B -  \nabla \times (u \times B) + \nabla \times ( (\nabla \times B) \times B)  = \Delta B , \label{eq:dtB}
\end{eqnarray}
All the physical parameters are chosen to be 1 for the simplicity of the presentation except the parameter $\nu$, which allows us to distinguish between inviscid flow ($\nu=0$) and viscous flow ($\nu>0$). The resistivity term (the right-hand side of (\ref{eq:dtB})) is essential for the well-posedness. The following theorem is the first step of the present paper. It shows the existence of global weak solutions in the whole space case. 

\begin{theorem}
Let $\nu > 0$ and  $u_0, B_0 \in L^{2} (\Bbb R^3)$,
with $\nabla \cdot u_0=0$. Then, there
exists a global weak solution $u, B \in L^{\infty}(\Bbb R_{+}; L^{2} (\Bbb R^3)) \cap L^{2} (\Bbb R_{+} ;
H^1 (\Bbb R^3))$ satisfying energy inequality
\begin{eqnarray*}
&&\hspace{-1cm}
\frac12 \left(  \| u(\cdot,t) \|_{L^2}^{2} +  \| B(\cdot,
t)\|_{L^2}^{2} \right)
+ \nu \int_0 ^t  \| \nabla   u(\cdot s) \|_{L^2}^{2}ds
+  \int_0 ^t \| \nabla  B(\cdot, s)\|_{L^2}^{2}ds  \\
&&\hspace{8cm}
\le \frac12 \left(  \| u_0 \|_{L^2}^{2} +  \| B_0\|_{L^2}^{2}  \right)
\end{eqnarray*}
for almost every $t\in [0, \infty)$.  Furthermore, if $\nabla \cdot
B_0=0$, then we have $\nabla \cdot B(\cdot, t)=0$ for all $t>0$.
\label{thm:global_weak}
\end{theorem}

\noindent
A previous version of this theorem in the case of a periodic domain has been proved in \cite{Acheritogaray_etal_KRM11} using a Galerkin approximation. Here, in the whole space case, the proof is based on mollifiers (see eqs. (\ref{eq:dtu_mol})-(\ref{eq:dtB_mol})) and the main estimates will be given at Proposition \ref{prop:energy} below. We note that this theorem does not require that $\nabla \cdot B_0 = 0$. However, if $\nabla \cdot B_0 = 0$, the divergence free condition is propagated.

The main results of this paper are the establishment of short-time existence of smooth solutions and a blow-up criterion (Theorem \ref{thm:blowup}). We also establish the existence of global smooth solutions for small data (Theorem \ref{thm:global_small}). Additionally, we show the uniqueness of solutions (Theorem \ref{thm:unique}). Finally, we state a Liouville theorem for smooth stationary solutions. 

For the sharp blow-up criterion, we need to introduce the following functional setting. We recall the homogeneous Besov space $\dot{B}^0_{\infty, \infty}$, which is
defined as follows. Let $\{\psi_k\}_{k\in \Bbb Z}$ be the
Littlewood-Paley partition of unity, where the Fourier transform
$\hat{\psi}_k (\xi)$ is supported on the annulus $\{ \xi \in \Bbb
R^N\, |\, 2^{k-1}\leq |\xi|< 2^{k}\}$(see e.g. \cite{Chemin_Clarendon98, Triebel_Birkhauser93}). Then,
$$
f\in \dot{B}^0_{\infty, \infty}\quad \mbox{if and
  only if }\quad \sup_{k\in \Bbb Z} \|\psi_k *f\|_{L^\infty} =:
  \|f\|_{\dot{B}^0_{\infty,
\infty}}<\infty.
$$
The following is a well-known embedding result, c.f. \cite[pp. 244]{Triebel_Birkhauser93}
 \bb L^\infty(\Bbb
R^N)\hookrightarrow \mbox{BMO}(\Bbb R^N)
\hookrightarrow\dot{B}^0_{\infty, \infty} (\Bbb R^N), \ee
where $\mbox{BMO}$ denotes the Bounded Mean Oscillation space \cite{Triebel_Birkhauser93}. Now, we state the theorems. The first one is the local existence theorem for smooth solutions and the blow-up criterion:

\begin{theorem}
Let $m>5/2$ be an integer, $\nu \geq 0$ and  $u_0, B_0 \in H^m (\Bbb R^3)$
with $\nabla \cdot u_0=0$. Then:
\begin{itemize}
\item[(i)] There
exists $T=T(\|u_0\|_{H^m}, \|B_0\|_{H^m})$ such that there exists a
unique solution $u, B \in L^{\infty}([0, T); H^m (\Bbb R^3))\cap Lip (0, T;
H^{m-2} (\Bbb R^3))$.
\item[(ii)] Define
\begin{equation}
X(t):=1+\|B(\cdot, t)\|_{H^m} ^2 + \| u(\cdot, t)\|_{H^m}^2,
\label{eq:defX}
\end{equation}
and
\begin{equation}
A(t):= \|\omega(t)\|_{\dot{B}^0_{\infty, \infty}} + \frac{1+\|u(t)\|_{L^\infty} ^2+\|B(t)\|_{L^\infty}^2+\|\nabla B(t)\|_{L^\infty}^2}{1+\log (1+ \|u(t)\|_{L^\infty}+\|B(t)\|_{L^\infty}+\|\nabla B(t)\|_{L^\infty})  }
\label{eq:defC}
\end{equation}
where we denoted $\omega=\nabla \times u$ the vorticity.
For $T^*< \infty$ then the following two statements are equivalent:
\begin{eqnarray}
&(i)& \quad X(t) < \infty, \quad \forall t < T^* \quad \mbox{ and } \quad \limsup_{t \to T^*} X(t) = \infty,
\label{eq:X(t)toinfty} \\
&(ii)& \quad \int_0^{t} A(s) \, ds < \infty, \quad \forall t < T^* \quad \mbox{ and } \quad \int_0^{T^*} A(s) \, ds = \infty.
\label{eq:intC(t)infty}
\end{eqnarray}
If such $T^*$ exists, then, $T^*$ is called the first-time blow up and (\ref{eq:intC(t)infty}) is a blow-up criterion.
\end{itemize}
\label{thm:blowup}
\end{theorem}

\noindent Note that this theorem is valid in both the viscous ($\nu>0$) and inviscid ($\nu=0$) cases. Next, we state the global existence theorem for smooth solutions with small data: 

\medskip
\begin{theorem}
Let $m>5/2$  be an integer, $\nu >0$, and $u_0, B_0 \in H^m (\Bbb R^3)$ with $\nabla \cdot u_0=0$.
There exists a universal constant $K=K(m, \nu)$ such
that if $\|u_0\|_{H^m} +\|B_0 \|_{H^m} < K$, then, there exists a
unique solution $ u, B \in L^{\infty}(\Bbb R_{+}; H^m (\Bbb R^3))\cap Lip
(\Bbb R_{+}; H^{m-2} (\Bbb R^3)) $.
\label{thm:global_small}
\end{theorem}

\noindent This theorem is only valid in the viscous case ($\nu>0$). The next theorem states the uniqueness of the solution: 

\medskip
\begin{theorem}
Let $(u_1, B_1)$ and $(u_2, B_2)$ are two week solutions.
\begin{itemize}
\item Assume $\nu \ge 0$, and $(u_2, B_2)$ satisfies
$$
   \int_0 ^{T}
   \left(
   \|\nabla u_2 \|_{L^\infty} +\|u_2 \|_{L^\infty} ^2+\|B_2 \|_{L^\infty}^2+\|\nabla B_2 \|_{L^\infty}^2
   \right) dt < \infty,
$$
then we have $u_1\equiv u_2$, and  $B_1\equiv B_2$ a.e., in $(0,T)\times \Bbb R^{3}$.

\item Assume $\nu >0$,
$u_2 \in L^{\infty} \big(0, T; H^1 (\Bbb R^3)\big) \cap L^{2} \big(0, T; H^2 (\Bbb R^3)\big)$,
$B_2 \in L^{2} \big(0, T; W^{1,\infty} (\Bbb R^3)\big)$,
Then we have $u_1\equiv u_2$, and  $B_1\equiv B_2$ a.e., in $(0,T)\times \Bbb R^{3}$.
\end{itemize}
\label{thm:unique}
\end{theorem}

\noindent More precisely, this theorem states that, if two solutions exist with the same data and if one of them is smooth, then they must coincide. The first statement is valid in both the viscous ($\nu>0$) and inviscid ($\nu=0$) cases but requires stronger regularity on the smooth solution. The second result is only valid in the viscous case ($\nu>0$) and the regularity for the $u$-component of the smooth solution reduces to that of the strong solution.

Finally, we state a Liouville type theorem for the smooth solutions of the following stationary Hall-MHD system.
\begin{eqnarray}
& & \hspace{-1cm}    u \cdot \nabla u + \nabla p  =  (\nabla \times
B) \times B + \nu\Delta u,
\label{ueq} \\
& & \hspace{-1cm} \nabla \cdot u = 0 , \label{divu1}\\
& & \hspace{-1cm}-  \nabla \times (u \times B) + \nabla \times (
(\nabla \times B) \times B)  = \Delta B , \label{beq}\\
& & \hspace{-1cm} \nabla \cdot B = 0 . \label{divb1}
\end{eqnarray}
The Liouville theorem reads as follows:

\begin{theorem}
 Let $(u, B) $
 be a $C^2(\Bbb R^3)$ solution to (\ref{ueq})-(\ref{divb1}) satisfying
 \bb
 \int_{\Bbb R^3} |\nabla u|^2dx +\int_{\Bbb R^3} |\nabla B|^2 dx
 <\infty,
 \ee
 and
 \bb
 u, B \in L^\infty (\Bbb R^3)\cap L^{\frac{9}{2}} (\Bbb R^3).
 \ee
We assume $\nu >0$. Then, we have $u=B=0$.
\label{thm:Liouville}
\end{theorem}

\begin{remark} If we set $B=0$ in the Hall-MHD system, the
above theorem reduces to the well-known Galdi result \cite{Galdi_Springer94} for the Navier-Stokes equations.
\label{rem:Galdi}
\end{remark}

\setcounter{equation}{0}
\section{Proofs of the main results}
\label{sec:proof}

We use the mollifier technique as described in \cite{Majda_Bertozzi_Cambridge01}. We consider the following mollifier operator:
$$
  {\mathcal J}_{\varepsilon} v = \rho_{\varepsilon} * v, \qquad  \rho_{\varepsilon} = \varepsilon^{-3} \rho(x/\varepsilon),
$$
where $\rho$ is a non-negative $C_0^\infty$ function, with unit integral. We introduce the regularized system as follows:
\begin{eqnarray}
& & \hspace{-1cm}  \partial_t u_\varepsilon
+   {\mathcal J}_{\varepsilon} \big( ({\mathcal J}_{\varepsilon} u_\varepsilon \cdot \nabla)  {\mathcal J}_{\varepsilon} u_\varepsilon \big)
+ \nabla p_\varepsilon
=   {\mathcal J}_{\varepsilon} \big( (\nabla \times  {\mathcal J}_{\varepsilon} B_\varepsilon) \times  {\mathcal J}_{\varepsilon} B_\varepsilon \big)
+ \nu \Delta {\mathcal J}^{2}_{\varepsilon} u_\varepsilon,
\label{eq:dtu_mol} \\
& & \hspace{-1cm} \nabla \cdot u_\varepsilon = 0 ,
\label{eq:divu_mol}\\
& & \hspace{-1cm} \partial_t B_\varepsilon
-  \nabla \times  \Big({\mathcal J}_{\varepsilon} \big( {\mathcal J}_{\varepsilon} u_\varepsilon \times  {\mathcal J}_{\varepsilon} B_\varepsilon \big) \Big)
+ \nabla \times  \Big( {\mathcal J}_{\varepsilon}\big( (\nabla \times  {\mathcal J}_{\varepsilon} B_\varepsilon) \times  {\mathcal J}_{\varepsilon} B_\varepsilon \big)\Big)
 = \Delta  {\mathcal J}_{\varepsilon}^{2}B_\varepsilon ,
 \label{eq:dtB_mol}
\end{eqnarray}
with initial condition
$$
(u_\varepsilon, B_\varepsilon) \big|_{{t=0}}={\mathcal J}_{\varepsilon} (u_{0}, B_{0}).
$$

\noindent
First, we have the

\begin{proposition}
Let $m$ be an integer such that $m>5/2$, let $\varepsilon>0$, $\nu \geq 0$ and
$u_{0}, B_{0} \in H^m (\Bbb R^3)$,
with $\nabla \cdot u_{0}=0$. Then, there
exists a unique global solution $u_{\varepsilon}, B_{\varepsilon} \in C^{\infty}\big(\Bbb R_{+}; C^{\infty}\cap H^{m} (\Bbb R^3)\big)$ which satisfies:
\begin{itemize}
\item[(i)] Energy inequality:
\bq && \frac12 \left(  \| u_{\varepsilon}(\cdot, t)\|_{L^2}^{2} +
\| B_{\varepsilon} (\cdot, t)\|_{L^2}^{2}  \right)
  + \nu \int_0 ^t \| \nabla  {\mathcal J}_{\varepsilon}  u_{\varepsilon}(\cdot,
  s)\|_{L^2}^{2}ds
  + \int_0^t \| \nabla  {\mathcal J}_{\varepsilon}  B_{\varepsilon}(\cdot, s)\|_{L^2}^{2} ds\n \\
  && \leq
   \frac12 \left(  \| u_{0}\|_{L^2}^{2} +  \|
B_{0} \|_{L^2}^{2}  \right) \label{eq:energy} \quad \forall t\in (0,
\infty).\eq
\item[(ii)]
There are positive constants C depending only on $m$, and constant
$T$ depending only on $m$, $\|u_{0}\|_{H^m}$, and $\|B_{0}\|_{H^m}$ such
that:
\begin{equation}
 \left\| (u_{\varepsilon}, B_{\varepsilon} ) \right\|_{L^\infty (0,T; H^m({\Bbb R}^3) ) \cap Lip ( 0,T; H^{m-2}({\Bbb R}^3))}
   \le C(\|u_{0}\|_{H^m}^{2} + \|B_{0}\|_{H^m}^{2}) .
\label{eq:estimate:Hm}
\end{equation}
\end{itemize}
\label{prop:energy}
\end{proposition}

\noindent
{\bf Sketch of proof.} The existence and uniqueness comes directly from the abstract Picard iteration theorem in $H^m$ (see details in \cite{Majda_Bertozzi_Cambridge01}). The energy estimate comes directly from (\ref{eq:dtu_mol}), (\ref{eq:dtB_mol}). The proof of estimate (\ref{eq:estimate:Hm}) is almost identical as the proof of the a priori estimate (\ref{eq:estim_2}) of Theorem \ref{thm:blowup} below and is skipped. \endproof

\noindent
We can now prove Theorem \ref{thm:global_weak}.

\medskip
\noindent
{\bf Sketch of proof of Theorem \ref{thm:global_weak}.} Thanks to Proposition \ref{prop:energy}, we can construct a sequence of regularized solutions to problem (\ref{eq:dtu}), (\ref{eq:dtB}) and the energy estimate (\ref{eq:energy}) provides the required compactness allowing to pass to the limit $\varepsilon \to 0$ in the weak formulation and to get a global weak solution. The details of the functional analysis are the same as in \cite{Acheritogaray_etal_KRM11}. \endproof

\noindent
Property (ii) of Proposition \ref{prop:energy} will be used as an a priori estimate for the construction of regularized solutions in the proof of Theorem \ref{thm:blowup} below.

\begin{proposition}
Let $m>5/2$ be an integer.
Let $(u,B)$ be a smooth solution to (\ref{eq:dtu})-(\ref{eq:dtB}). Then, there are two positive universal constants $C_1$ and $C_2$ such that the following a priori estimates hold:
\begin{eqnarray}
&&\hspace{-1cm}
\frac{d}{dt}
(\|B\|_{H^m} ^2 + \|u\|_{H^m}^2 )+\|\nabla B\|_{H^m} ^2
+2\nu \|D u\|_{H^m}^2 \leq \nonumber \\
&&\hspace{0cm}
C_1 (1 + \|B\|_{L^\infty} ^2 +\|\nabla B\|_{L^\infty} ^2 + \|u\|_{L^\infty}^2 +\|\nabla
u\|_{L^\infty}) (\|B\|_{H^m} ^2 + \|u\|_{H^m}^2 +1) .
\label{eq:estimate_2} \\
&&\hspace{-1cm}
\frac{d}{dt}
(\|B\|_{H^m} ^2 + \|u\|_{H^m}^2 )+ 2 \|\nabla B\|_{H^m} ^2
+2\nu \|D u\|_{H^m}^2 \leq \nonumber \\
&&\hspace{0cm}
\leq C_2 \big(\|\nabla B\|_{H^m} ^2
+ \|\nabla u\|_{H^m}^2   \big) (\|B\|_{H^m}
^2 + \|u\|_{H^m}^2 +  \|u\|_{H^m} + \|B\|_{H^m} )  .
\label{eq:estimate_3}
\end{eqnarray}
\label{prop:estimate2}
\end{proposition}

\noindent
{\bf Proof.} We first concentrate ourselves on (\ref{eq:estimate_2}). Let $\alpha =(\alpha_1,\alpha_2, \alpha_3) \in {\mathbb N}^3$ be a multi-index. We operate $D^\alpha = \partial^{|\alpha|}/\partial x_1^{\alpha_1} \ldots \partial x_3^{\alpha_3}$ (where $|\alpha| = \alpha_1 + \ldots + \alpha_3$) on
(\ref{eq:dtu}) and (\ref{eq:dtB}) respectively and take the scalar product of them with $D^\alpha B$ and $D^\alpha u$ respectively, add them together and then sum the result over $|\alpha|\leq m$. We obtain
\begin{eqnarray}
&&\hspace{-1cm}
\frac12 \frac{d}{dt}(\| u\|_{H^m}^2+ \|B\|_{H^m}^2)
+\nu \|D u\|_{H^m}^2 + \|\nabla B\|_{H^m } ^2
= \nonumber\\
&&\hspace{4cm}
- \sum_{0<|\alpha|\leq m} \int_{\Bbb R^3} D^\alpha \big((\nabla \times B) \times B\big)\cdot D^\alpha (\nabla\times B) \, dx
\nonumber \\
&& \hspace{4cm}
+ \sum_{0<|\alpha|\leq m} \int_{\Bbb R^3} D^\alpha (u\times
B)\cdot D^\alpha (\nabla \times B )\, dx
\nonumber \\
&& \hspace{4cm}
-\sum_{0<|\alpha|\leq m} \int_{\Bbb R^3} D^\alpha (u\cdot \nabla u) \cdot D^\alpha u\,
dx
\nonumber \\
&& \hspace{4cm}
+\sum_{0<|\alpha|\leq m} \int_{\Bbb R^3} D^\alpha\{(\nabla \times B)\times
 B\} \cdot D^\alpha u \, dx
\nonumber \\
&& \hspace{4cm}
=: I_1+I_2 + I_3 +I_4.
\label{eq:I1-I4}
\end{eqnarray}
Notice that the $|\alpha|=0$ terms on the right hand side above have exactly cancelled each other by energy conservation. The cancellation is crucially important for the existence of global smooth solution for small data. Then, we estimate successively each of the $I_1$ to $I_4$ terms. We have:
\begin{eqnarray*}
I_1&=&
- \sum_{0<|\alpha|\leq m} \int_{\Bbb R^3} \left[ D^\alpha \big((\nabla \times B) \times B\big) -   (D^\alpha (\nabla \times B)) \times B   \right] \cdot D^\alpha (\nabla\times B) \, dx
\end{eqnarray*}
where the second term of the right-hand side is simply zero. Using the well-known calculus inequality,
\begin{equation}
\sum_{|\alpha|\leq m} \|D^\alpha (fg)-(D^\alpha f)g\|_{L^2} \leq C
(\|f\|_{H^{m-1}} \|\nabla g\|_{L^\infty} +\|f\|_{L^\infty} \|
g\|_{H^m}),
\label{eq:calc_id}
\end{equation}
we get:
\begin{eqnarray}
I_1
&\leq&  C(\|B\|_{H^m} \| \nabla B \|_{L^\infty} + \|\nabla B \|_{L^\infty} \|B\|_{H^m})
\|\nabla B\|_{H^{m}}
\label{eq:estimate_I1_2} \\
&\leq& \frac14\|\nabla B\|_{H^m}^2 +C \|B\|_{H^m}^2 \|\nabla
B\|_{L^\infty}^2 ,
\label{eq:estimate_I1_1}
\end{eqnarray}
On the other hand, using Leibnitz formula and the Sobolev inequality, we obtain
\begin{eqnarray}
I_2 &\leq &  \sum_{0<|\alpha|\leq m} \|D^\alpha (u\times B)\|_{L^2} \|\nabla B \|_{H^m} \nonumber \\
&\leq& C (\|u\|_{L^\infty} \|B\|_{H^m} + \| u\|_{H^m}
\|B\|_{L^\infty})\|\nabla B\|_{H^m} \nonumber \\
&\leq&  \frac14 \|\nabla B\|_{H^m}^2 +C\|u\|_{L^\infty}^2
\|B\|_{H^m}^2+  C\|u\|_{H^{m}} ^2 \|B\|_{L^\infty}^2.
\label{eq:I2}
\end{eqnarray}
Then, we remark that
\begin{eqnarray*}
I_3 &=&
-\sum_{0<|\alpha|\leq m} \int_{\Bbb R^3} [ D^\alpha (u\cdot \nabla u)- u\cdot \nabla D^\alpha u ] \cdot D^\alpha u\,
dx.
\end{eqnarray*}
Indeed, the second term is zero by the fact that $u$ is divergence free. Then, similarly to the above calculation, using the calculus inequality (\ref{eq:calc_id}), we obtain
\begin{eqnarray}
\hspace{-1cm}
I_3 &\leq&\sum_{0<|\alpha|\leq m} \|D^\alpha (u\cdot \nabla u)- u\cdot \nabla D^\alpha u
\|_{L^2}\|\nabla u\|_{H^{m-1}} \leq C \|\nabla u\|_{L^\infty} \|
\nabla u\|_{H^{m-1}}^2.
\label{eq:estimate_I3}
\end{eqnarray}
From (\ref{eq:I1-I4}) we get
\begin{eqnarray*}
I_4&\leq& \sum_{0<|\alpha|\leq m} \|(\nabla \times B)\times B\|_{H^m} \|\nabla u\|_{H^{m-1}}.
\end{eqnarray*}
Note that we can take $\|\nabla u\|_{H^{m-1}}$ instead of $\| u\|_{H^{m}}$ because $|\alpha|>0$. This remark is important for the proof of the next theorem about the global existence of smooth solutions for small data. Using Leibnitz formula, we derive
\begin{eqnarray}
 I_4
 &\leq & C (\|\nabla B\|_{L^\infty} \|B\|_{H^m} + \|\nabla B\|_{H^m} \|B\|_{L^\infty} ) \|\nabla u\|_{H^{m-1}} \label{eq:estimate_I4_2}
\\
 &\leq&C \|\nabla B\|_{L^\infty} \|B\|_{H^m} \|\nabla u\|_{H^{m-1}}
 +\frac12 \|\nabla B\|_{H^m}^2 + C \|B\|_{L^\infty} ^2\|u
 \|_{H^m}^2.
\label{eq:estimate_I4_1}
\end{eqnarray}
From estimates (\ref{eq:estimate_I1_1}), (\ref{eq:I2}), (\ref{eq:estimate_I3}) and (\ref{eq:estimate_I4_1}), we obtain
\begin{eqnarray}
&&\hspace{-1cm}
\frac{d}{dt}
(\|B\|_{H^m} ^2 + \|u\|_{H^m}^2 )+\|\nabla B\|_{H^m} ^2
+2\nu \|D u\|_{H^m}^2 \n \nonumber \\
&&\hspace{1cm}
\leq C(\|B\|_{L^\infty}^2 +\|\nabla B\|_{L^\infty}^2
+\|u\|_{L^\infty}^2 ) (\|B\|_{H^m}
^2 + \|u\|_{H^m}^2 ) \nonumber\\
&&\hspace{2cm}
 + C\|\nabla u\|_{L^\infty} \| \nabla u\|_{H^{m-1}}^2 + C \|\nabla B\|_{L^\infty} \|B\|_{H^m} \|\nabla u\|_{H^{m-1}},
\label{master}
\eq
from which we easily deduce (\ref{eq:estimate_2}).

We now turn towards estimate (\ref{eq:estimate_3}). It is deduced through a small change in the estimate (\ref{eq:I2}) for $I_2$. From Leibnitz formula, we have:
\begin{eqnarray}
I_2 &\leq &  \sum_{0<|\alpha|\leq m} \|D^\alpha (u\times B)\|_{L^2} \|\nabla B \|_{H^m} \nonumber \\
&\leq &  \sum_{0<|\alpha|\leq m} \sum_{j=1}^3 \|D^{\alpha-e_j} (\partial_{x_j} u\times B +  u\times \partial_{x_j}B)\|_{L^2} \|\nabla B \|_{H^m} \nonumber \\
&\leq& C (\|u\|_{L^\infty} \|\nabla B\|_{H^{m-1}} + \|\nabla u\|_{L^\infty} \|B\|_{H^{m-1}}  \nonumber \\
& & \hspace{2cm} + \|B\|_{L^\infty} \|\nabla u \|_{H^{m-1}} + \|\nabla B\|_{L^\infty} \|u\|_{H^{m-1}} )  \,  \|\nabla B\|_{H^m}
\nonumber \\
&\leq&  C ( \| u \|_{H^m} \|\nabla B \|_{H^{m-1}} + \| B \|_{H^m} \|\nabla u \|_{H^{m-1}} ) \,  \|\nabla B\|_{H^m}
\label{eq:I2_diff}
\end{eqnarray}
From estimates (\ref{eq:estimate_I1_2}), (\ref{eq:I2_diff}), (\ref{eq:estimate_I3}) and (\ref{eq:estimate_I4_2}), we easily deduce (\ref{eq:estimate_3}). This completes the proof of Proposition \ref{prop:estimate2}. \endproof

In order to prove Theorem \ref{thm:global_small}, we need to use the following

\begin{lemma}
Assume that $a$ is a positive constant, $x(t)$, $y(t)$ are two nonnegative $C^1({\mathbb  R}_+)$ functions, and $D(t)$ is a nonnegative function, satisfying
$$
\frac{d}{dt} \left( x^{2} + y^{2}\right) + D \le a  (x^{2} + y^{2} + x+ y) D.
$$
If additionally, the initial data satisfy
\begin{equation}
 x^{2}(0) + y^{2}(0) + \sqrt{ 2(x^{2}(0) + y^{2}(0)) }  <  \frac{1}{a},
\label{eq:ini_cond}
\end{equation}
then, for any $t>0$, one has
$$
  x^{2} (t)+ y^{2}(t) + x(t) + y(t) <  x^{2}(0) + y^{2}(0) + \sqrt{2( x^{2}(0) + y^{2}(0))} < \frac1a
$$
\label{lem:lyapounov}
\end{lemma}

\medskip
\noindent
{\bf Proof:}
Notice that
$$
\frac{d}{dt} \left( x^{2} + y^{2}\right) + D \le a  (\, x^{2} + y^{2} + \sqrt{2(x^{2} + y^{2})} \,) D
$$
Since (\ref{eq:ini_cond}) is true initially, it is still true for short time, so that one has
$$
x^{2}(t) + y^{2}(t) + \sqrt{2(x^{2}(t) + y^{2}(t))} < \frac{1}{a}
$$
Then $ x^{2}(t) + y^{2}(t)$ is a decreasing function in this short period. Hence
$$
 x^{2}(t) + y^{2}(t) + \sqrt{2( x^{2}(t) + y^{2}(t)) }  \le    x^{2}(0) + y^{2}(0) + \sqrt{2(x^{2}(0) + y^{2}(0))}
$$
Then by an extension argument, it holds true for all time. \endproof

\medskip
\noindent
{\bf Proof of Theorem \ref{thm:blowup}.}
We construct a sequence of weak solutions of the regularized system (\ref{eq:dtu_mol})-(\ref{eq:dtB_mol}) and remark that such solutions do actually satisfy the a priori estimate (\ref{eq:estimate_2}). This allows us to pass to the limit in a subsequence and show the existence of smooth solutions on short times.

Below we set
$$X(t):=\|B(\cdot, t)\|_{H^m} ^2 + \| u(\cdot, t)\|_{H^m}^2
+1. $$
Then, from (\ref{eq:estimate_2}), and using the Sobolev inequality,
we have
 \bqn
  \frac{d}{dt} X &\leq& C (1 + \|B\|_{L^\infty} ^2 +\|\nabla B\|_{L^\infty} ^2 + \|u\|_{L^\infty}^2 +\|\nabla
u\|_{L^\infty})X \n \\
&\leq& C (1 + \|B\|_{H^m}^2  +\| u\|_{H^m}^2 +\| u\|_{H^m} ) X\n \\
&\leq & C X^2. \eqn
Therefore, thanks to nonlinear Gronwall's inequality, we have:
$$ X(t)\leq \frac{X(0)}{1-C_0 X(0) t}. $$
Now, choose $T=\frac{1}{ 2C_0 X(0)}$. Then:
\bb
\label{loc}
X(t)\leq 2 X(0) \quad \forall t\in [0,
T).
\ee
This implies the following a priori estimate:
$$ \left\| (u, B )\right\|_{L^\infty (0,T; H^m({\Bbb R}^3) )}
   \le C(\|u_{0}\|_{m}^{2} + \|B_{0}\|_{m}^{2}). $$
Now, thanks to a direct estimate on the time derivatives using eqs. (\ref{eq:dtu}), (\ref{eq:dtB}), we have also the a priori estimate:
$$ \left\| (u, B )\right\|_{ Lip ( 0,T; H^{m-2}({\Bbb R}^3))}
  \le C(\|u_{0}\|_{m}^{2} + \|B_{0}\|_{m}^{2}). $$
Adding these two estimates together, we get the a priori estimate:
\begin{equation}
\left\| (u, B ) \right\|_{L^\infty (0,T; H^m({\Bbb R}^3)) \cap Lip( 0,T; H^{m-2}({\Bbb R}^3))}   \le C(\|u_{0}\|_{m}^{2} + \|B_{0}\|_{m}^{2}).
\label{eq:estim_2}
\end{equation}
Estimate (\ref{eq:estim_2}) also holds true for the modified equations (\ref{eq:dtu_mol})-(\ref{eq:dtB_mol}). This is exactly estimate (\ref{eq:estimate:Hm}) of Proposition \ref{prop:energy}.
As in the proof in Theorem \ref{thm:global_weak}, there exists a subsequence, still denoted by $(u_\varepsilon, B_\varepsilon)$ whose limit gives rise to a global weak solution $(u,B)$. Using the lower-semicontinuity of the norm, one has that this weak solution satisfies the estimate
$$
 \left\| (u, B ) \right\|_{L^\infty (0,T; H^m({\Bbb R}^3)) \cap Lip ( 0,T; H^{m-2}({\Bbb R}^3))}
   \le C(\|u_{0}\|_{m}^{2} + \|B_{0}\|_{m}^{2}),
$$
which proves the local existence of a smooth solution on $[0,T)$. Uniqueness follows from the same kind of proof as Theorem \ref{thm:unique}.

Next, in order to prove the blow-up criterion (\ref{eq:intC(t)infty}) we recall
the following version of Beale-Kato-Majda type logarithmic Sobolev inequality in $\Bbb
R^3$(see \cite[formula 14.2]{Taylor_AMS00} or \cite{Chemin_Clarendon98}),
 \bb\label{log}
  \|f\|_{L^\infty} \leq C(1+\|f\|_{\dot{B}^0_{\infty, \infty}}
  )\{\log (1+ \|f\|_{H^{m-1}} )\},\quad m >5/2 .
  \ee
Substituting $f$ for $\nabla u$ in (\ref{log}) and inserting the result into (\ref{master}), we
have
\begin{eqnarray}
&&\hspace{-1cm}
\frac{d}{dt} X \leq  C(1+\|B\|_{L^\infty}^2 +\|\nabla B\|_{L^\infty}^2
+\|u\|_{L^\infty}^2)X \nonumber \\
&&\hspace{3cm}
+ C (1 +  \|\nabla u\|_{\dot{B}^0_{\infty, \infty}})  X
\log(1+X) . \label{eq:dtX}
\end{eqnarray}
We use the fact that the Calderon-Zygmund operator is bounded from the homogeneous Besov space ${\dot{B}^0_{\infty, \infty}}$ into itself \cite{Triebel_Birkhauser93}, namely we have
\begin{eqnarray}
&&\hspace{-1cm}
\|\nabla u\|_{\dot{B}^0_{\infty, \infty}} \le C \|\omega\|_{\dot{B}^0_{\infty, \infty}}.
\label{eq:CZ}
\end{eqnarray}
We also use the Sobolev inequality to obtain
\begin{eqnarray}
&&\hspace{-1cm}
  1+\|B\|_{L^\infty}^2 +\|\nabla B\|_{L^\infty}^2
+\|u\|_{L^\infty}^2 \nonumber \\
&&\hspace{2cm}
\le C \frac{1+\|B\|_{L^\infty}^2 +\|\nabla B\|_{L^\infty}^2
+\|u\|_{L^\infty}^2}{1+\log (1+ \|u\|_{L^\infty}+\|B\|_{L^\infty}+\|\nabla B\|_{L^\infty})  } \log(1+ X) .
\label{eq:Sob}
\end{eqnarray}
Inserting (\ref{eq:CZ}) and (\ref{eq:Sob}) into (\ref{eq:dtX}), we get:
 \bqn
\frac{d}{dt} X &\leq &C \, A(t) \, X \ln (1+X), \eqn
where $A(t)$ is given by (\ref{eq:defC}). We now recall that by Sobolev imbedding, there exists $C>0$ such that we have $A(t) \leq C X(t)$. Then, by this inequality and Gronwall's lemma, we obtain the equivalence of  (\ref{eq:X(t)toinfty}) and (\ref{eq:intC(t)infty}). \endproof

\begin{remark}
By applying the same argument, we can show the local well-posedness for the simple Hall problem without coupling to the fluid velocity $u$, which is written as follows:
$$
\partial_t B  + \nabla \times ( (\nabla \times B) \times B)  = \Delta B ,
$$
with a initial data $B_{0}$.
We can also extend the result to the following generalized Hall problem
$$
\partial_t B  + \nabla \times ( (\Lambda^{\alpha} \nabla \times B) \times B)  = -  \Lambda^{\beta} B ,
$$
where $\Lambda^{\alpha}$ is the fractional power of the Laplacian $\Lambda^{\alpha} = (-\Delta)^{\alpha/2}$, supplemented
with a initial data $B_{0}$.  With the same argument, we can show that
if $\beta \ge \alpha +2$, $\alpha \ge 0$, then this generalized Hall problem is also locally well posed. When $\beta < \alpha +2$
it is an open problem to determine if this problem is well-posed or not.
\label{rem:extension}
\end{remark}

\medskip
\noindent
{\bf Proof of Theorem \ref{thm:global_small}.}
We use the inequality (\ref{eq:estimate_3}), and estimate, using the Sobolev inequality,
\begin{eqnarray}
&&\hspace{-1cm}
\frac{d}{dt}
(\|B\|_{H^m} ^2 + \|u\|_{H^m}^2 )+ 2 \|\nabla B\|_{H^m} ^2
+2\nu \|\nabla u\|_{H^m}^2 \nonumber \\
&&\hspace{0cm}
\leq C(\|B\|_{L^\infty}^2 +\|\nabla B\|_{L^\infty}^2
+\|u\|_{L^\infty}^2  ) (\|B\|_{H^m}
^2 + \|u\|_{H^m}^2 )  \nonumber \\
&&\hspace{1cm}
+ C\|\nabla u\|_{L^\infty} \|\nabla u\|_{H^{m-1}}^2 + C \|\nabla B\|_{L^\infty} \|B\|_{H^m}  \|\nabla u\|_{H^{m-1}} \nonumber \\
&&\hspace{0cm}
\leq C_1 \big(\|\nabla B\|_{H^m} ^2
+ \|\nabla u\|_{H^m}^2   \big) (\|B\|_{H^m}
^2 + \|u\|_{H^m}^2 +  \|u\|_{H^m} + \|B\|_{H^m} )  .
\label{master_2}
\end{eqnarray}
Therefore if
 $$
  \|B_0\|_{H^m} ^2 + \|u_0\|_{H^m}^2 + \sqrt{  \|B_0\|_{H^m} ^2 + \|u_0\|_{H^m}^2 }  <  \frac{\min\{ 1, 2\nu\}}{C_1},
$$
then, by Lemma \ref{lem:lyapounov},  we have for any $t>0$
  \bb
\frac{d}{dt} (\|B(\cdot, t)\|_{H^m} ^2 + \|u (\cdot, t) \|_{H^m}^2 ) \leq 0,
\ee
and
\bqn
\|B(t)\|_{H^m} ^2 + \|u(t)\|_{H^m}^2 \le \|B_0\|_{H^m} ^2 + \|u_0\|_{H^m}^2 \leq \frac{\min\{ 1, 2\nu\}}{C_1} ,
  \eqn
for all $t > 0$. Hence the so obtained solution is global in time, which ends the proof. \endproof

\medskip
\noindent
{\bf Proof of Theorem \ref{thm:unique}.}
The proof of this theorem uses the same estimates as for the proof of Theorem \ref{thm:blowup} (i) applied to $u=u_1 - u_2$ and $B = B_1 - B_2$. The details are omitted. \endproof

\medskip
\noindent
{\bf Proof of Theorem 2.5 } We first estimate the pressure in
(\ref{ueq}). Taking the divergence of (\ref{ueq}), and using the
identity, $ (\nabla \times B)\times B=-\nabla \frac{|B|^2}{2}
+(B\cdot \nabla) B,$ we obtain,
 $$
  \Delta\left( p+\frac{|B|^2}{2}\right)=-\sum_{j,k=1}^3
  \partial_j\partial_k (u_ju_k -B_jB_k),
  $$
  from which we have the representation formula of the pressure,
  using the Riesz transforms in $\Bbb R^3$,
  \bb
 p=\sum_{j,k=1}^3 R_jR_j (u_ju_k-B_jB_k) -\frac{|B|^2}{2}.
 \ee
 By the Calderon-Zygmund inequality, one has
 \bb\label{cd}
 \|p\|_{L^q}\leq C\|u\|_{L^{2q}}^2 +C\|B\|_{L^{2q}}^2,\quad\
 1<q<\infty,
 \ee
 if $u,B\in L^{2q} (\Bbb R^3)$. Let $\si$ be the standard
cut-off function defined as follows. Consider $\sigma\in C_0
^\infty(\Bbb R^N)$ such that
 $$
   \sigma(|x|)=\left\{ \aligned
                  &1 \quad\mbox{if $|x|<1$},\\
                     &0 \quad\mbox{if $|x|>2$},
                      \endaligned \right.
$$
and $0\leq \sigma  (x)\leq 1$ for $1<|x|<2$.  Then, for each $R
>0$, let us define
$$
\s \left(\frac{|x|}{R}\right):=\s_R (|x|)\in C_0 ^\infty (\Bbb R^N).
$$
We take the inner product of Eq. (\ref{ueq}) with $u\si$ and the inner product of Eq. (\ref{beq}) with $B \si$, add the result together and integrate over ${\mathbb R}^3$. After integration by parts, we have
\bq\label{lio}
 \lefteqn{\nu\int_{\Bbb R^3} |\nabla u|^2 \si dx +\int_{\Bbb R^3} |\nabla
 B|^2\si dx
 =\frac12 \int_{\Bbb R^3} |u|^2 (u\cdot \nabla )\si dx +\int_{\Bbb
 R^3} p (u\cdot \nabla )\si dx} \n \\
 &&\qquad -\int_{\Bbb R^3} u\times B\cdot \nabla \si \times B
 dx +\int_{\Bbb R^3} (\nabla \times B)\times B\cdot \nabla \si
 \times B dx \n \\
 &&\qquad +\frac{\nu}{2}\int_{\Bbb R^3} |u|^2
 \Delta \si dx +\frac{1}{2}\int_{\Bbb R^3} |B|^2
 \Delta \si dx\n \\
 &&\quad\qquad :=I_1+\cdots +I_6.
 \eq
 We have the following estimates,
\bqn |I_1| &\leq& \int_{\{R\leq |x|\leq 2R\}}|u|^3 |\nabla \si|dx\\
&\leq& \frac{1}{2R} \|\nabla \s\|_{L^\infty} \left(\int_{\{R\leq
|x|\leq 2R\}} |u|^{\frac{9}{2}} dx\right)^{\frac23}
\left( \int_{\{R\leq |x|\leq 2R\}} \, dx\right)^{\frac13}\n \\
&\leq& C \|u\|_{L^{\frac92} (R\leq |x|\leq 2R)}^3 \to 0 \quad\mbox{
as $R\to \infty$}. \eqn
  Using the estimate (\ref{cd}), one has
\bqn |I_2| &\leq&\int_{\{R\leq |x|\leq 2R\}} |p||u| |\nabla \si|dx\\
&\leq&\frac{1}{R} \|\nabla \s\|_{L^\infty} \left(\int_{\Bbb R^3}
|p|^{\frac{9}{4}} dx\right)^{\frac49}\left(\int_{\{R\leq |x|\leq
2R\}} |u|^{\frac92} dx\right)^{\frac29}
\left( \int_{\{R\leq |x|\leq 2R\}} \, dx\right)^{\frac13}\n \\
&\leq &C(\|u\|_{L^{\frac92}}^2 +\|B\|_{L^{\frac92}}^2)
\|u\|_{L^{\frac92} (R\leq |x|\leq 2R)}\to 0 \quad\mbox{ as $R\to
\infty$}. \eqn
 \bqn |I_3| &\leq& \int_{\{R\leq |x|\leq 2R\}}  |u| |B|^2 |\nabla \s| dx\n \\
&\leq& \frac{1}{R} \|\nabla \s\|_{L^\infty} \left(\int_{\Bbb R^3}
|u|^{\frac{9}{2}} dx\right)^{\frac29}\left(\int_{\Bbb R^3}
|B|^{\frac92} dx\right)^{\frac49}
\left( \int_{\{R\leq |x|\leq 2R\}} \, dx\right)^{\frac13}\n \\
&\leq &C \|B\|_{L^{\frac92}}^2 \|u\|_{L^{\frac92} (R\leq |x|\leq
2R)}\to 0 \quad\mbox{ as $R\to \infty$}. \eqn

\bqn|I_4| &\leq& \int_{\{R\leq |x|\leq 2R\}}  |\nabla B| |B|^2 |\nabla \s| dx\n \\
&\leq& \frac{1}{R} \|\nabla \s\|_{L^\infty} \|B\|_{L^\infty}\|\nabla
B\|_{L^2} \left(\int_{\{R\leq |x|\leq 2R\}} |B|^{6}
dx\right)^{\frac16}
\left( \int_{\{R\leq |x|\leq 2R\}} \, dx\right)^{\frac13}\n \\
&\leq & C\|B\|_{L^\infty}\|\nabla B\|_{L^2} \|B\|_{L^{6} (R\leq
|x|\leq 2R)}\to 0 \quad\mbox{ as $R\to \infty$}, \eqn
 since $\|B\|_{L^6}\leq C \|\nabla B\|_{L^2} <\infty$ by the Sobolev
 embedding. Then,
\bqn
 |I_5| +|I_6|&\leq& C \int_{\{R\leq |x|\leq 2R\}} (|u|^2 +|B|^2) |\Delta \s| dx\n \\
&\leq& \frac{C}{R^2} \|D^2\s\|_{L^\infty}\left(\int_{\{R\leq |x|\leq
2R\}}( |u|^2 +|B|^2)^3 dx\right)^{\frac13}
\left( \int_{\{R\leq |x|\leq 2R\}} \, dx\right)^{\frac23}\n \\
&\leq & C (\|u\|_{L^6 (R\leq |x|\leq 2R)}^2 +\|B\|_{L^6 (R\leq
|x|\leq 2R)}^2)\to 0 \quad\mbox{ as $R\to \infty$}. \eqn
 Therefore, passing to the limit $R\to \infty$ in (\ref{lio}), and using the
 dominated convergence theorem, one has
 $$\int_{\Bbb R^3} |\nabla u|^2 dx +\int_{\Bbb R^3} |\nabla B|^2
 dx=0,
 $$
which implies the conclusion of the theorem. \endproof


\end{document}